a

# Boron Content and the Superconducting Critical Temperature of Carbon-Based Materials

(November 26, 2020)

Nadina Gheorghiu, Charles R. Ebbing, and Timothy J. Haugan

*Abstract*—In this paper, we present results on magnetization properties of boron nitride-carbon (BN-C) and boron carbide-carbon ($B_4$C-C) granular mixtures. The temperature-dependent magnetization for field-cooled during cooling and field-cooled during warming shows a kind of thermal hysteresis that is always seen around a metamagnetic phase transition from an antiferromagnetic martensite to a ferromagnetic austenite phase. The low-temperature magnetization has an upward turn that can be attributed to superparamagnetism, diamagnetic shielding, and trapped flux characteristic to high-temperature superconducting materials. After subtracting the diamagnetic background, the field-dependent magnetization loops $M(B)$ are ferromagnetic-like, more significant for the BN-C than for the $B_4$C-C mixture. In addition, the magnetization loops show the kink feature characteristic to granular superconductivity. The irreversibility temperature for a $B_4$C-C mixture having 37.5 wt% B is $T_c \cong 76$ K. Combining our data with previous results on B-doped diamond and Q-carbon, we find that $T_c$ increases linearly with the B concentration.

I. INTRODUCTION

Power systems, including for aerospace applications, need to optimally combine advanced functions with low-cost production. Currently, most of the high-temperature superconducting materials (HTS) materials are rare-earth based. The more available, thus cost-effective carbon (C) allotropes and C-based materials are ultra-high-strength yet much lighter than copper or steel. These materials have showed magnetic, superconducting (SC), and even HTS properties. Crucially, the power loss is expected to be significantly reduced in C-based HTS composites. In particular, boron(B)-doped C composites showed increasingly higher critical temperatures $T_c$. B-doped diamond is SC below $T_c \cong 2.8$ K [1]. Nanocrystalline diamond grown on BN is SC below $T_c \cong 3.4$ K [2]. Nanocrystalline diamond B-doped at 3.7% is SC under $T_c \cong 5$ K [3]. Recently, (B)-doped Q-carbon (B-QC) composites have showed increasingly higher $T_c$ as the B concentration was increased [4]. Calculations also found that $T_c$ for heavily B-doped C materials could reach 60 K [5] or even 80 K [6].

In this research, we have probed magnetic properties of B-doped C granular mixtures. The temperature-dependent magnetization for field-cooled during cooling (FCC) and field-cooled during warming (FCW) shows thermal hysteresis, which is observed around a metamagnetic phase transition from the antiferromagnetic (AFM) martensite to the ferromagnetic (FM) austenite phase. The low-temperature magnetization has an upward turn as a manifestation of superparamagnetism, diamagnetic (DM) shielding, and trapped flux characteristic to HTS materials. Subtraction of the DM background from the magnetization loops $M(B)$ reveals FM, which appears to be more significant for the BN-C than for the $B_4$C-C mixture. In addition, the magnetization loops show the kink feature characteristic to granular SC. At the highest ever reported 37.5 wt% B content, the temperature where the zero-field cooled (ZFC) curve departs from the FCC curve also known as the irreversibility temperature is $T_c \cong 76$ K. Combining our three data points with previously reported results on SC in B-diamond and B-QC, we find a linear relationship between $T_c$ and the B concentration.

This work was supported by The Air Force Office of Scientific Research (AFOSR) for the LRIR #14RQ08COR & LRIR #18RQCOR100 and Aerospace Systems Directorate (AFRL/RQ). N. Gheorghiu acknowledges Dr. G.Y. Panasyuk for discussions and for his continuous support and inspiration.
N. Gheorghiu was with UES, Inc., Dayton, OH 45432, USA. (Nadina.Gheorghiu@yahoo.com).
C.R. Ebbing is with University of Dayton Research Institute, Dayton, OH 45469, USA.
T.J. Haugan is with the Air Force Research Laboratory (AFRL), Aerospace Systems Directorate, AFRL/RQ, Wright-Patterson AFB, OH 45433, USA.

## II. EXPERIMENT

The 99%+ C graphite powder with a grain size 7 μm (RS Mines, Sri Lanka) and either BN or $B_4C$ (Alfa Aesar) micron-size flakes were carefully milled using the pestle-and-mortar technique to uniform mixtures with a B content 10 wt%, 22 wt%, and 37.5 wt%, respectively. Fig. 1 shows the Scanning Electron Microscopy data for the 22 wt% B content. Temperature- and field-dependent magnetization measurements were obtained using the 6500 Quantum Design Physical Properties Measurement System in the temperature range 1.9 - 300 K and for magnetic fields of induction up to 1 T.

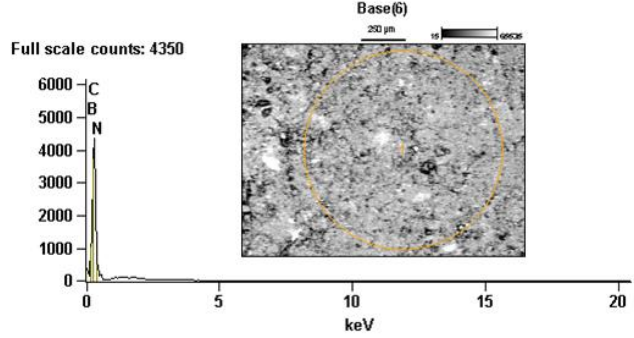

Fig. 1. SEM for a BN-C mixture having a B = 22% content. The bright spots are BN flakes.

## III. RESULTS AND DISCUSSION

### A. *Temperature-dependent magnetization measurements*

The materials were first investigated for their temperature-dependent magnetic properties. During the FCC process, the Meissner flux expulsion occurs under the influence of flux pinning and the ZFC curve usually goes under the FCC curve for temperatures below the irreversibility temperature $T_{irrev}$. Temperature-dependent magnetization plots for a BN-C mixture of mass 14.1 mg with a B content 22 wt% under a magnetic field of strength $H = 2$ kOe are shown in Fig. 2. We see that the Néel point (at the PM to AFM transition) remains the same, $T_N \cong 58$-$60$ K. The three curves separate at the same temperature of about 68 K. The observed difference between the FCC and FCW branches reflects the existence of a thermal hysteresis in the sample, which is always observed around a metamagnetic phase transition from the AFM martensite to the FM austenite phase. Thus, it appears that FM and AFM clusters are formed within the sample. In C/graphite nanoflakes, there are AFM correlations between unlike sublattices (*ABAB*...) and FM correlations between like sublattices (*AAA*... or *BBB*...). The ZFC and FCC curves intersect two times: at the irreversibility temperature $T_{irrev} \cong 48$ K and at $T_v \cong 68$ K. The irreversibility temperature $T_{irrev}$ is the merging point of the ZFC and FCC (or FCC and FCW), not of the ZFC and FCW magnetization curves [7]. The FCW and FCC curves practically coincide below $T_{irrev}$ and above $T_v$. Above $T_v$, all three curves coincide, i.e., there is no thermal hysteresis. Increasing the field strength to $H = 3$ kOe brings the ZFC curve above the FCC curve (Fig. 3). This unusual result of having the ZFC magnetic moment larger than the one at FCC was also found for nitrogen(N)-doped diamond and LSMO/YBCO bilayers [8], as well as in SC phosphorus-doped disordered graphite [9] and graphite [10]. The source for this behavior is the weaker FM-superparamagnetic component for the FCC than for the ZFC magnetization, as it relates to stronger SC (Meissner) contribution to the ZFC magnetic moment than to the FCC one.

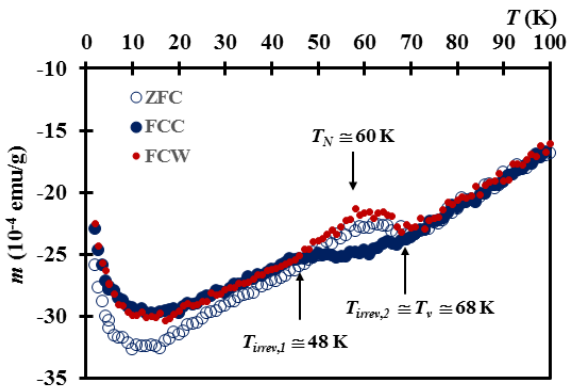

Fig. 2. ZFC, FCC, and FCW magnetization plots under a field strength $H = 2$ kOe for a BN-C mixture with the B content 22%.

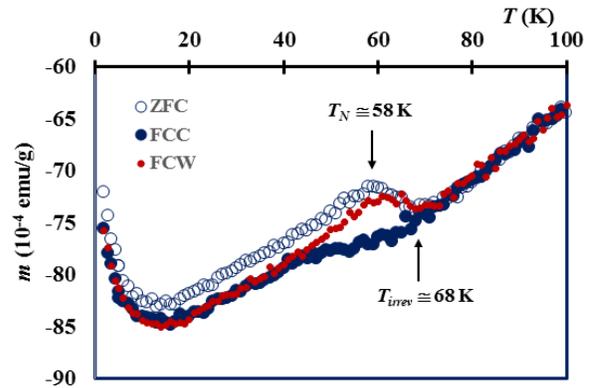

Fig. 3. ZFC, FCC, and FCW magnetization plots under a field strength $H = 3$ kOe for a BN-C sample with a B content 22 wt%.

New features were revealed in the temperature-dependent magnetization plots for a $B_4C$-C mixture with the B content 37.5 wt% (Fig. 4). The sample's mass was 13.6 mg. There is no Néel transition in this case. Significantly, $T_c \cong 76$ K is close to the predicted maximum value of 80 K at 38 wt% B [6] and it also suggests a possible diamond-like D05 B-C crystal structure [11]. Using the high-temperature data, we find for the FM contribution $\chi_{ferro} = C/(T - T_{Curie})$ both the Curie constant and corresponding transition temperature: $C \cong 136\times10^{-4}$ K·emu/g and $T_{Curie} \cong 107$ K for ZFC, $C \cong 140\times10^{-4}$ K·emu/g and $T_{Curie} \cong 104$ K for FCC, respectively. The temperature-dependent magnetization data was replotted after the subtraction of the FM component (Fig. 5). While the difference between the FCC and ZFC magnetizations is positive below $T_c$ as it would be in a SC state, the two magnetizations are indistinguishable when the FM component is excluded. This result suggest that it is precisely the FM component that is responsible for the SC-like anisotropy in the magnetic moment seen in the departing of the ZFC curve below the FCC curve for $T \leq T_c$. While B-rich carbides are FM [12], herein we find that a $B_4C$-C mixture has features of a ferromagnetic-superconductor. Obviously, the only way to accommodate both SC and FM would be through a triplet spin state. We also notice that in the $B_4C$-C mixture, the soft layers of the semimetal graphite (a metal at thickness layer larger than 50 nm [13]) are sandwiched between the hard and insulating $B_4C$ flakes such that Cooper pairing of electrons from the metallic layers [14] is entirely possible.

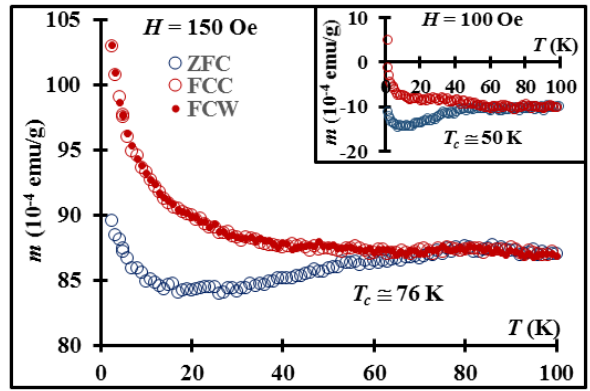

Fig. 4. ZFC, FCC, and FCW in a field $H = 150$ Oe for a $B_4C$-C mixture with the B content 37.5%. The inset shows data on an undoped graphite sample.

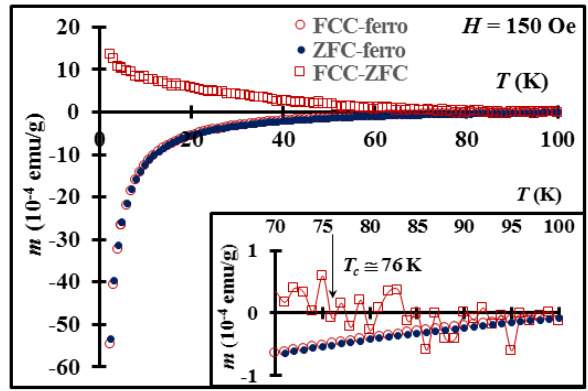

Fig. 5. FCC and ZFC magnetization without the FM contribution, also the difference FCC-ZFC in a field strength $H = 150$ Oe for a $B_4C$-C mixture with the B content 37.5 wt%.

B. *Field-dependent magnetization measurements*

Field-dependent magnetization loops $m(B)$ with and without the DM background were obtained for magnetic induction up to 1 T and for selected temperatures from 300 K to 1.9 K. The sample's DM overcomes its FM only at the lowest temperature. After removing the DM baseline, the BN-C mixture with the B content 22 wt% shows FM at all temperatures. By comparison, the $B_4C$-C mixture with the B content 37.5 wt% shows much smaller magnetization for the same field interval. In addition, while the $m(B)$ loops appear FM before and after the subtraction of the DM background, there are in fact significantly more DM and more hysteretic when compared to the 22 wt% B case. At $T = 60$ K $< T_c$ and without the DM background, the $m(B)$ loops have the SC-like shape. In addition, the $m(B)$ hysteresis loops show the kink that is specific to granular SC.

Thus, the higher wt% B content produced a mixture with a higher $T_c$. We then added our results to other published data on B-doped C (Table I) and plotted is as $T_c$(wt% B) in Fig. 8. The line's slope is 2.1 K/(wt% B) with a square residual 99%. The $T_c$(wt% B) line's intercept is at the critical value (for SC) $x_c \cong 0.52$ wt% B concentration. Notice that $x_c$ is smaller than the 2.8 wt% B content that made diamond SC [1].

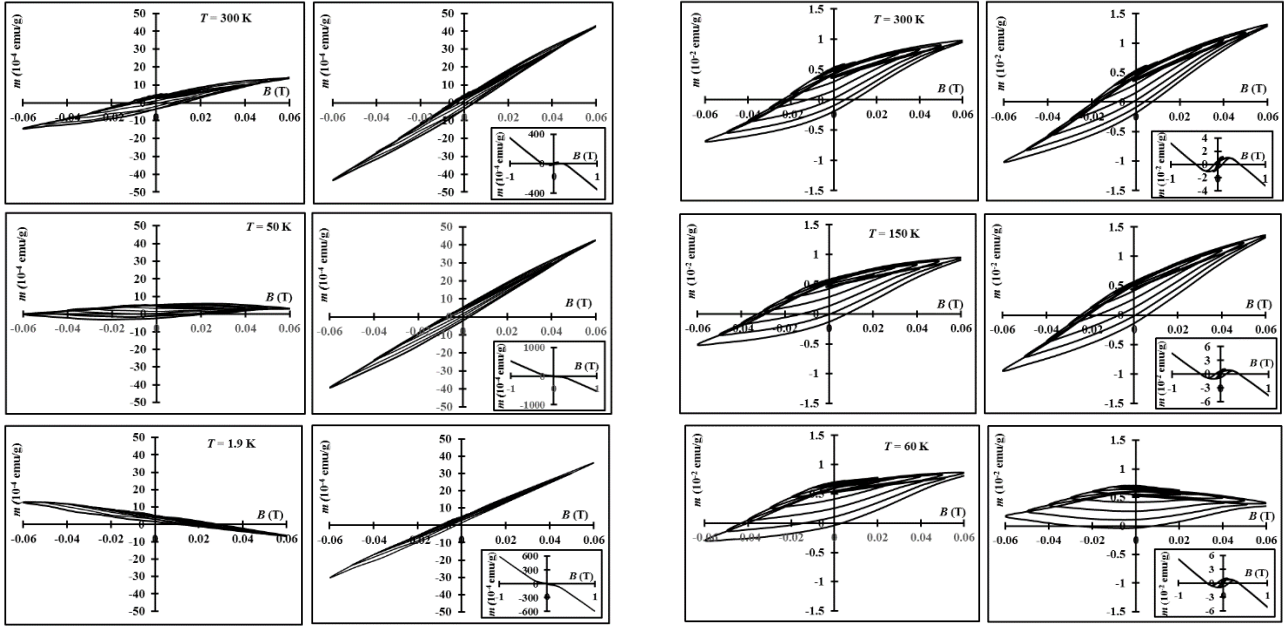

Fig. 6. Hysteretic magnetization loops for a BN-C mixture with the B content 22%, with (left) and without DM background (right).

Fig. 7. Hysteretic magnetization loops for a B$_4$C-C mixture with the B content 37.5 wt%, with (left) and without DM background (right).

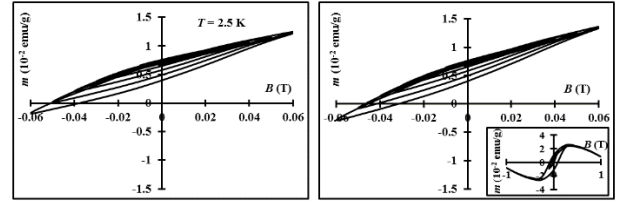

TABLE I
DEPENDENCE OF THE CRITICAL TEMPERATURE ON THE BORON CONTENT

| B(wt%) | $T_c$(K) | Reference |
|---|---|---|
| 2.8 | 4 | E.A. Ekimov et al., Nature **428**, 542 (2004) |
| 3.7 | 5 | K. Ishizaka et al., PRL **100**, 16602 (2008) |
| 17 | 37 | A. Bhaumik et al., ACS Nano **11**, 5351 (2017) |
| 22 | 48 | This work |
| 25 | 55 | A. Bhaumik et al., ACS Nano **11**, 11915 (2017) |
| 37.5 | 76 | This work |

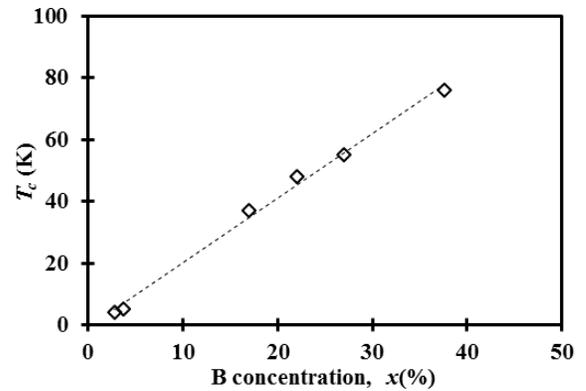

Fig. 8. The critical temperature is linear in the B(wt%) content.

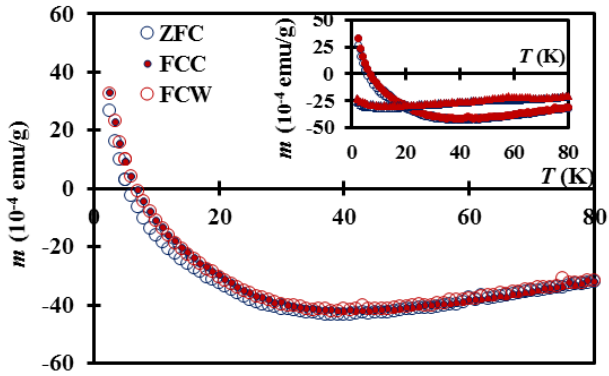

Fig. 9. ZFC-FCC-FCW magnetization plots for a BN-C mixture with the B content 10 wt% at a field strength $H = 2$ kOe. The inset also shows data for the 22 wt% B mixture. The two sets of curves cross at $T \cong 19$ K, vs. a critical temperature of 20 K for a 10% B content as predicted from the linear dependence $T_c$(wt% B) (Fig. 8).

While another BN-C (mass 11.1 mg) with a smaller B content of 10 wt% did not show significant thermal hysteresis and clearly no martensitic transition (Fig. 9), the magnetization curves cross the ones for the 22 wt% B mixture at $T \cong 19$ K. By comparison, the $T_c$(wt% B) line from Fig. 8 would give a $T_c \cong 20$ K for a 10 wt% B content.

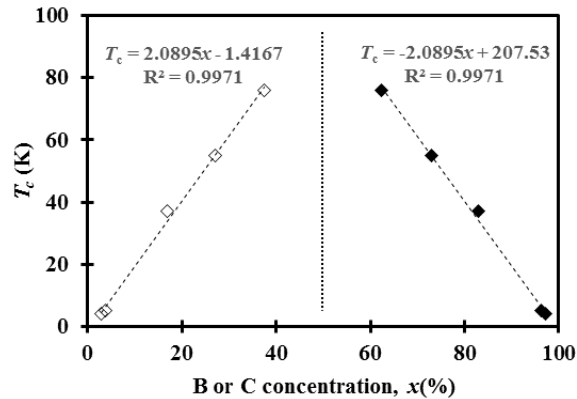

Fig. 10. Perfect symmetrical $T_c(x)$ dependences for the B(wt%)-C and C(wt%)-B mixtures. Hole doping of C by B (empty symbols) vs. electron doping of B by C (filled symbols). The B-line intercepts at $x_c \cong 0.68\%$ (where $T_c = 0$). The C-line intercepts at $T_c \cong 208$ K (where $x = 0\%$), while the B-line reaches the same $T_c \cong 208$ K at $x = 100\%$.

Electronically, the B⁻ sheets are graphite-like sheets. One can also find a perfect symmetry between the hole doping of C by B and the electron doping of B by C, as illustrated in Fig. 9. The data point for the BN-C was not included. The positive slope B-line (hole doping) and the negative slope C-line (electron doping) intersect at an equal (50%-50%) B-C mixture with a $T_c \cong 103$ K. This happens to be the same as the Curie temperature found before. A $T_c$ larger than 100 K was predicted for B-doped Q-C [15]. In our previous work, both transport and magnetization SC-like behavior were found below $T_c \sim 50$ K in oxygen-implanted graphites, including diamond-like C films [16]. The linear $T_c$(wt% B) relationship found here suggests that a $T_c \sim 50$ K can be achieved in a B-C system at an either 25 wt% hole or a 25 wt% electron concentration. In fact, we have found $T_c \sim 50$ K in undoped graphite foil (mass 7.8 mg, see inset in Fig. 4). Such a system contains many graphite layers twisted relative to each other, thus being an excellent host for both interface SC with flat-band energy bands and disordered-induced magnetism. Moreover, excitonic SC involving the formation of electron-hole pairs [17], observed below 50 K in the hydrogenated graphites [18,19], might even lead to room-temperature SC. For graphene stacks, theoretical $T_c$ vs. the chemical potential $\mu$ displays a dome-shaped curve, just like in 122 pnictides and cuprate superconductors [20]. Calculations shows that B-doped fully hydrogenated multilayer graphene (graphene) can reach $T_c \sim 150$ K [21].

IV.  CONCLUSION

Two-dimensional C-based materials show a myriad of interesting physical properties such as magnetism and SC [22]. Each of the materials used in this work have been either proven or predicted to be SC: graphite [23], twisted bilayers of graphene [24] or BN [25], and $B_4C$ [26]. It should not come as a surprise that SC can also be found in certain mixtures prepared from these materials. In this work, we have found that B-doping of C can also be achieved by using B-rich materials like BN and $B_4C$. The $B_4C$-C mixture with a B content of 37.5 wt%, which is the highest so far reported B-doping of a C allotrope, showed FM and an irreversibility temperature $T_c \cong 76$ K. Adding to our data previous results on B-doped diamond or B-doped Q-carbon, we have found that the critical temperature $T_c$ increases linearly with the B concentration. Moreover, the magnetization loops show the kink feature characteristic to granular SC. In addition, temperature-dependent magnetization measurements show that the BN-C mixture goes through a metamagnetic phase transition from an AFM martensite to a FM austenite phase.